\documentstyle[12pt]{article}
\renewcommand{\baselinestretch}{1.65}

\begin{document}

\renewcommand{\baselinestretch}{1.1}

\title{\Large \bf Topology and Fragility in Cosmology  
                                        \vspace{3mm} \\} 

\author{
M. J. Rebou\c{c}as\thanks{
{\sc internet: reboucas@cat.cbpf.br}  }, \ \ 
R. K. Tavakol\thanks{
{\sc internet: reza@maths.qmw.ac.uk}  }   \ \   and   \ \
A. F. F. Teixeira\thanks{ 
{\sc internet: teixeira@novell.cat.cbpf.br} } \\
\\  
\vspace{1mm}
$^{\ast}~^{\ddagger}~$Centro Brasileiro de Pesquisas F\'\i sicas \\
                      Departamento de Relatividade e Part\'\i culas \\
                      Rua Dr.\ Xavier Sigaud 150 \\
                      22290-180 Rio de Janeiro -- RJ, Brazil \\
\\
$~^{\dagger}~$School of Mathematical Sciences  \\
              Queen Mary and Westfield College \\
              Mile End Road \\
              London E1 4NS -- England \vspace{2mm} \\
        }        

\date{\today} 

\maketitle

\vspace{2mm}       
%%%%%%%%%%%%%%%%%%%%%%%%%%%%%%%%%%%%%%%%%%%%%%%%%%%%%%%%%%%%
\begin{abstract} \vspace{3mm} 
%%%%%%%%%%%%%%%%%%%%%%%%%%%%%%%%%%%%%%%%%%%%%%%%%%%%%%%%%%%%%
We introduce the notion of topological fragility and
briefly discuss some examples from the literature.
An important example of this type of fragility is the way
globally anisotropic Bianchi~V generalisations of the 
FLRW $k=-1$ model result in a radical restriction on
the allowed topology of spatial sections, thereby
excluding compact cosmological models with
negatively curved three-sections with anisotropy.
An outcome of this is to exclude chaotic mixing in 
such models, which may be relevant, given the many recent
attempts at employing compact FLRW $k=-1$ models 
to produce chaotic mixing in the cosmic microwave background 
radiation, if the Universe turns out to be globally anisotropic.
\end{abstract}

{\bf key words:} cosmology, topology, chaotic mixing, fragility. 

\vspace{3mm}
                       
%%%%%%%%%%%%%%%%%%%%%%%%%%%%%%%%%%%%%%%%%%%%%%%%%%%%%%%%%%%%%%%%%%%
\section{Introduction}
%%%%%%%%%%%%%%%%%%%%%%%%%%%%%%%%%%%%%%%%%%%%%%%%%%%%%%%%%%%%%%%%%%

It is well known that general relativity (GR) is a {\em local\/} 
metrical theory and therefore the corresponding Einstein field 
equations do not fix the {\em global\/} topology of spacetime. 
Given this freedom in the topology of the spacetime 
manifold \footnote{Here, in line with its usage in the literature,
by the {\em topology of spacetime\/} we mean the topology of
the $t=const$ spacelike section ${\cal M}_3$ of the spacetime 
manifold ${\cal M}_4$.}, 
a question arises as to how free the choice of these 
topologies may be and how one may hope to determine them,
which in turn is intimately related to the observational 
consequences of the spacetime possessing non-trivial topologies.

These questions have motivated two sets of work: (i)
those attempting to tabulate mathematically the set of 
possible topologies for the spacetime,
given certain symmetry constraints, such as homogeneity%
~\cite{Weeks1985}~--~\cite{FujiwaraIshiharaKodama1993}, 
and (ii) those relating to the possible observational 
(or physical) consequences of adopting particular 
topologies for the spacetime%
~\cite{Gott1980}~--~\cite{LachiezeReyLuminet1995}. 

To determine the actual topology of the spacetime, 
one would have to ultimately rely on the
{\em observations--dynamics--topology correspondences\/},
in the sense of looking at those observational or dynamical 
features of the universe (or cosmological models) which are 
dependent on the topology of spacetime.
These correspondences can take various forms, 
such as for example the existence of (i)   
dynamical solutions with physically identifiable properties
which can only arise with certain choices of topology~%
\cite{OliveiraReboucasTeixeira1994}, (ii) identifiable images
of galaxies implying closed 3-spaces~% 
\cite{Gott1980,EllisSchreiber86,Fagundes1993}, and 
(iii) features, such as potential mixing of the cosmic 
microwave background radiation (CMWBR), which could possibly 
be identified, through the studies of the corresponding spectra, 
as being induced by non-trivial topology of the spacetime%
~\cite{Lochart1982,Gurzadyan1992,EllisTavakol1994,%
CornishSpergelStarkman1996a}.

To use these correspondences effectively, however, 
it is important to study their nature, by examining 
whether they are {\em sensitive} in the sense that
(i) changes in spacetime topology produce observable 
dynamical consequences, and (or)
(ii) changes in the assumptions underlying cosmological models
(such as symmetry) can have severe constraining effects 
on the allowed spacetime topologies. 

We shall refer to such sensitivity as 
{\em topological fragility} \footnote{As a counterpart to
the concept of {\em dynamical fragility} introduced 
elsewhere~\cite{TavakolEllis1988,ColeyTavakol1992}.}.
If present, such fragility could have important consequences:
it could either facilitate or hinder the 
task of inferring the topology of 
the spacetime, depending upon its precise nature. 

Our aim in this work is to point out, with the help of examples, t
hat such topological fragilities can arise naturally in usual 
cosmological modelling and could therefore be consequential 
in practice.
\vspace{3mm}
%%%%%%%%%%%%%%%%%%%%%%%%%%%%%%%%%%%%%%%%%%%%%%%%%%%%%%%%%%%%%%%
\section{Topology and geometry in cosmology}
%%%%%%%%%%%%%%%%%%%%%%%%%%%%%%%%%%%%%%%%%%%%%%%%%%%%%%%%%%%%%%%

Even though the metrical structure of a space does not 
generally fix its topology, the geometry can in 
certain settings severely constrain the possible set of 
allowed topologies. 
For example, in the case of compact 2-manifolds, there is a 
well known  relationship between the topology and the geometry~%
\cite{Weeks1985,Scott1983}.
However, in the case of 3-manifolds, ${\cal M}_3$, the situation 
is much more involved. In particular, for a general spacetime
geometry very little can be said about the underlying topology of 
the spacetime manifold ${\cal M}_4$. 

General asymmetric spacetime geometries are, however, rarely
the object of study in cosmological modelling. What is usually
done in practice is to impose simplifying assumptions (such as those
involving symmetry) in order to reduce the resulting field equations 
to a manageable form. 
In particular, to reduce the complicated nonlinear set of 
Einstein's partial differential equations to a manageable 
set of ordinary differential equations, it is usual to assume 
spatial homogeneity~\cite{MacCallum1979}.

In what follows we shall assume that the spacetime manifold 
${\cal M}_4$ is decomposable into the form
${\cal M}_4 = {\cal R}  \times {\cal M}_3$, 
where the spacelike 3-manifolds ${\cal M}_3$ are orientable, 
connected and complete Riemannian manifolds. These are
the main topological properties one might expect in any 
reasonable model of the universe~\cite{Ellis1971}.
Further we shall assume the spacetime to be expanding
and homogeneous, which would include the 
Friedmann-Lema\^{\i}tre-Robertson-Walker (FLRW) and 
the Bianchi models.

We shall show in the following sections that changes in the 
symmetry properties of the universe can have significant 
constraining effects on the allowed topologies. In this sense
such models are topologically fragile.
\vspace{3mm}
%%%%%%%%%%%%%%%%%%%%%%%%%%%%%%%%%%%%%%%%%%%%%%%%%%%%%%%%%%%%
\section{The FLRW setting} 
%%%%%%%%%%%%%%%%%%%%%%%%%%%%%%%%%%%%%%%%%%%%%%%%%%%%%%%%%%%%

Standard cosmological models, almost universally employed
for the purpose of interpretation of observations, are the
spatially homogeneous and isotropic
Friedmann--Lema\^{\i}tre--Robertson--Walker (FLRW)
spacetime manifolds \footnote{With their flat, elliptic 
or hyperbolic constant spatial curvatures being specified 
by the curvature parameter $k = 0, \pm1$, respectively.}
${\cal M}_4$, which can be split into
${\cal R} \times {\cal M}_3$, possessing three-dimensional
spacelike $t=const$ hypersurfaces of homogeneity ${\cal M}_3$.

If these 3-surfaces are assumed to be {\em globally\/}
homogeneous-and-isotropic, i.e. to possess a continuous 
six-parameter isometry group acting transitively on the 
whole 3-spaces ${\cal M}_3$, then the correspondence 
between the geometry and topology of the 3-spaces is very tight
and  results in: 
${\cal R}^3$ (the Euclidean 3-space) for $k=0$, the 3-sphere 
${\cal S}^3$ and the projective 3-space ${\cal P}^3$ for the 
$k=+1$ case, and the hyperbolic 3-space ${\cal H}^3$ for the 
$k=-1$ case. 

The assumption of global homogeneity-and-isotropy of 3-spaces 
${\cal M}_3$ is, however, too restrictive and not necessarily 
demanded by cosmological observations%
~\cite{Ellis1971,DemianskiLapucha1987,LachiezeReyLuminet1995}.
As a result, it is customary to adopt a less
restrictive setting of {\em local\/} homogeneity-and-isotropy%
~\cite{Ellis1971,DemianskiLapucha1987,LachiezeReyLuminet1995}.

A word of clarification is in order here: corresponding to
each 3-manifold (${\cal M}_3,\,{\bf g}$), there exists a 
simply connected covering manifold 
(${\cal \widetilde{M}}_3 \:, {\bf \tilde{g} } $) 
such that        
(${\cal M}_3,\,{\bf g}$) is obtained from 
(${\cal \widetilde{M}}_3 \:, {\bf \tilde{g} }$)
by identifying points in ${\cal \widetilde{M}}_3 $ which are 
equivalent under a discrete group of isometries of 
${\cal \widetilde{M}}_3$. In other words, ${\cal M}_3 $ 
is obtained by  forming the quotient space 
${\cal M}_3 = {\cal \widetilde{M}}_3 / \Gamma$,
where $\Gamma$ is a discrete group of isometries 
of ${\cal \widetilde{M}}_3 $ without 
fixed points, acting properly discontinuously%
~\cite{Ellis1971}.
By construction (${\cal \widetilde{M}}_3 \:, {\bf \tilde{g} }$)
is {\em locally\/} indistinguishable from
(${\cal M}_3,\,{\bf g}$). The {\em global\/} 
features
\footnote{Such as, for example, whether or not there is 
unicity of geodesics between two fixed points.} 
can, however, be quite different.
The identification of points in ${\cal \widetilde{M}}_3$ 
via $\Gamma$ produces 3-manifolds ${\cal M}_3$ which are 
multi-connected, and usually admit a lower-dimensional
group of isometries. So for the FLRW cases, for example,
one usually obtains quotient manifolds 
${\cal M}_3 = {\cal \widetilde{M}}_3 / \Gamma$
which do not admit the full six-dimensional group of
isometries ($G_6$), i.e., the quotient manifolds
${\cal M}_3 $ are not maximally symmetric. 
This amounts to saying that one or more of the 
linearly independent Killing vector fields associated with the 
isotropies alone, and defined on ${\cal \widetilde{M}}_3$
by the FLRW metric, are excluded by the topological 
identification, since they cannot be globally defined 
on ${\cal M}_3$. And yet ${\cal M}_3$ is still 
{\em locally\/} homogeneous-and-isotropic. In other words,
the metric tensor of (${\cal M}_3, {\bf g} $) is the same at 
every point, but 
${\cal M}_3 = {\cal \widetilde{M}}_3 / \Gamma$ is
not globally isotropic since it does not permit global maximal 
symmetry.
It should be stressed that in general topological
identifications lower 
the dimension of the group of isotropies, breaking the 
global isotropy of the 3-spaces~%
\footnote{The only exception is the orientable
compact projective 3-space ${\cal P}^3$, whose covering
space is the 3-sphere ${\cal S}^3$. This follows because the 
isotropy group $H_p$ of any point $p \in {\cal S}^3$ leaves invariant
precisely that point and the antipodal point, which in
turn is the identification one uses to build the quotient
manifold ${\cal P}^3 \equiv {\cal S}^3 / \Gamma$.}.
Clearly, the breaking of the global isotropy is apparent
in many cases, since the identifications define preferred 
directions.

Now it is only with the assumption of {\em local\/} 
homogeneity-and-isotropy, that many other topological alternatives 
become possible for ${\cal M}_3$~\cite{Ellis1971,Wolf1967}. 
So, for example, in the case of FLRW $k=0$  there are six orientable
and compact 3-manifolds, whereas the $k=\pm 1$ cases 
allow an infinite number of orientable and compact
topological alternatives for the $t=const$ 3-manifolds.

As an example of topological fragility in the FLRW 
setting, we recall the way the assumption of global 
homogeneity-and-isotropy radically restricts the 
choice of possible topologies \footnote{Of course we are 
assuming that observationally (i.e. within the present 
observational range and accuracy) local and global 
homogeneities cannot be distinguished - i.e.\ up to this
level of resolution, observations are {\em stable} with 
respect to such changes. If this turns out not to be the case, 
then this will become an example of {\em resolution induced} 
topological fragility. This is interesting as it highlights 
how the nature of fragility might also depend on the scope 
and accuracy (and hence the epoch) of observations.}. 

As other examples of topological fragility we recall 
that one cannot have an isolated electric charge in any 
orientable compact 3-space (by using Gauss's law).    
This amounts to saying that the overall electric charge 
in such spaces is related to the topology, which in turn 
may explain one feature of the observed universe that 
would otherwise be an arbitrary initial 
condition~\cite{EllisSchreiber86}.
Further, it has recently been shown~%
\cite{OliveiraReboucasTeixeira1994} that a discrete 
change in the topology of ${\cal M}_3$ from the 
${\cal S}^3$ to the quaternionic manifold ${\cal Q}^3$ 
can exclude certain solutions of the Maxwell's equations. 
In this way, changes in topology can induce important 
dynamical (physical) consequences.
\vspace{3mm}
%%%%%%%%%%%%%%%%%%%%%%%%%%%%%%%%%%%%%%%%%%%%%%%%%%%%%%%%%%%%%
\section{Bianchi models and topology}
%%%%%%%%%%%%%%%%%%%%%%%%%%%%%%%%%%%%%%%%%%%%%%%%%%%%%%%%%%%%%

\begin{sloppypar}
In the FLRW setting, the complexity of the topological structures
and the set of alternatives contrast strikingly with the
simplicity of the local metric properties. The next more general 
setting usually considered in cosmological modelling is that of 
the homogeneous anisotropic Bianchi models.
Now the introduction of local anisotropy 
greatly reduces the richness of allowed topologies,
since in this case the so called space-form 
problem is simpler than in the FLRW case \footnote{For
the Kantowski-Sachs cases the covering space
is ${\cal S}^2 \times {\cal R}^1$. For all the other such
locally homogeneous Bianchi cases the covering space 
${\cal \widetilde{M}}_3$ has the topology ${\cal R}^3$, except for the
Bianchi type IX where ${\cal \widetilde{M}}_3 = {\cal S}^3$.},
which in turn is due to the fact that even though one still 
has similar discrete translations, there are fewer reflections 
and rotations which could be combined with these~\cite{Ellis1971}.
\end{sloppypar}

For the case of locally homogeneous Bianchi types, a 
correspondence between the eight Thurston types of homogeneous
geometries~\cite{Thurston1979,Thurston1982} and the Bianchi 
types can be set up. This has been recently discussed by Fagundes%
~\cite{Fagundes92} and Fujiwara {\em et al.\/}%
~\cite{FujiwaraIshiharaKodama1993}.  

An important outcome of the latter work% 
~\cite{FujiwaraIshiharaKodama1993} has been to show 
that no anisotropic expansion is allowed for the 
Bianchi~V model with a closed (compact without boundary) 
spatial section. 
This result, which we shall use in the next section,
can be understood from the following argument.
Let (${\cal M}_3,\,{\bf g}$) be a locally homogeneous
spacelike section of a Bianchi~V spacetime. According
to Milnor~\cite{Milnor1976}, (${\cal M}_3,\,{\bf g}$) is 
necessarily locally isometric to a maximally symmetric
3-space of negative constant curvature, i.e. it locally 
admits the hyperbolic geometry. On the other hand, 
Mostow's~\cite{Mostow1973} (see also~\cite{Thurston1982}) 
rigidity theorem ensures that if two closed hyperbolic 3-manifolds 
are homeomorphic then they are isometric. This amounts to
saying that Bianchi~V spacetimes are rigid, permitting 
only isotropic expansion, i.e. they allow only an overall 
change in the scale factor. 

\vspace{3mm}
%%%%%%%%%%%%%%%%%%%%%%%%%%%%%%%%%%%%%%%%%%%%%%%%%%%%%%%%%%%%%%
\section{Fragility of mixing in closed FLRW  ${\bf k = -1}$ 
          models}
%%%%%%%%%%%%%%%%%%%%%%%%%%%%%%%%%%%%%%%%%%%%%%%%%%%%%%%%%%%%%% 

Recently compact FLRW $k=-1$  models have been considered as
examples of relativistic cosmological models which possess
rigorous (chaotic) mixing properties~\cite{Lochart1982,Gurzadyan1992,%
EllisTavakol1994,HaywardTwamley1990,CornishSpergelStarkman1996a}.
{}From a general point of view, the geodesic flows on compact negative 
curvature manifolds have been known to result in $K$-flows%
~\cite{Sinai1960} with the corresponding Kolmogorov entropy 
given by 
\begin{equation} \label{sinaieq}
K \propto \frac{1}{V^{1/D}} \,,
\end{equation}
where $V$ is the volume of the closed manifold and $D$ is
its dimension~\cite{Sinai1960}. This clearly shows that 
$K \rightarrow 0$ as $V \rightarrow \infty$,  
as for example in the case of flows on open ${\cal H}^3$.
This, however, is an {\em all time} result which as it 
stands is not very informative for the case of the 
universe with a finite lifetime. In such cases, one can 
still derive useful information. In particular, it has been 
shown that for such models the deviation of neighbouring
geodesics is sensitively dependent on the cosmological 
density parameter $\Omega_0$ and the redshift $z$. For example 
the maximum distance apart of such geodesics initially 
making an angle $\alpha$ at the surface of decoupling 
$z=z_d$ is given by~\cite{EllisTavakol1994}
\begin{equation}  \label{maxdis}
\overline \delta(z_d) = \frac{\alpha R_0 \sinh (\,\lambda (z_d)\,)}
{1 + z_d} \,,
\end{equation}
where 
$\lambda (z)$ is an analytic function of $z$ and 
$\Omega_0$ given in eq.\ (10) of ref.~\cite{EllisTavakol1994}, 
and $R_0$ is the value of the scale parameter at $t = t_0$.
For significant mixing of null geodesics to occur one would 
require~\cite{EllisTavakol1994} 
\begin{equation} \label{mixcond} 
f = \frac{\overline \delta (z_d)}{L_c (t_d)} \alpha >> 1\,,
\end{equation}
where $L_c (t_d)$ is the topological compactification
scale calculated at $t = t_d$.  
In this way a measure of the effective mixing 
can be obtained once an estimate of 
$L_c$ and $\Omega_0$ is given.

Now despite the enormous success of FLRW models, they are 
nevertheless approximate, with the real universe unlikely
to be truly isotropic and homogeneous. The question then
arises as to whether such mixing can still occur if, 
for example, one of these symmetry restrictions is removed.
Here as a first step, we look at the effects of including 
anisotropies.

As natural anisotropic generalisation of the FLRW $k=-1$ 
isotropic models, we consider the Bianchi type~V anisotropic 
models which also possess negative curvature everywhere on 
their three-spatial sections. 
To see this explicitly, we may consider the example of the 
locally rotationally symmetric anisotropic Bianchi~V
model given by the metric~\cite{CollinsEllis1979}
\begin{equation}
ds^2 = dt^2 - a^2 (t) \: e^{2z}\, (dx^2 + dy^2)
- b^2 (t)\, dz^2 \,,
\end{equation}
where $a$ and $b$ are differentiable functions of $t$.
Clearly as $a \rightarrow b$ the model tends towards the
$k=-1$ FLRW isotropic model and hence it can be treated as 
its simplest anisotropic generalisation.
 
Now as was pointed out in the previous section,
FLRW $k=-1$ isotropic models can possess compact spatial 
three-surfaces and therefore give rise to mixing, as 
quantified by~(\ref{maxdis}) and~(\ref{mixcond}).
On the other hand, recalling that Bianchi~V homogeneous
compact universes allow only a change of the overall 
scale factor~\cite{FujiwaraIshiharaKodama1993}, it is 
immediately clear that the presence of non-zero
(Bianchi~V type) anisotropy destroys the possibility
of having closed space sections, which in
turn using (\ref{sinaieq}) inhibits the possibility 
of mixing in such models.

As a result global (Bianchi~V type) anisotropic expansion 
and closed space sections are not compatible and therefore the 
mixing property of the compact $k=-1$ models is not stable 
with respect to such anisotropic generalisations.
Consequently, smoothing of the microwave background radiation
by such a method becomes questionable.
\vspace{3mm}
%%%%%%%%%%%%%%%%%%%%%%%%%%%%%%%%%%%%%%%%%%%%%%%%%%%%%%%%%%
\section{Conclusion}
%%%%%%%%%%%%%%%%%%%%%%%%%%%%%%%%%%%%%%%%%%%%%%%%%%%%%%%%%%

We have introduced the concept of topological
fragility and pointed out cosmological examples 
where it is present.
Such fragility can be of potential significance,
as is demonstrated by the fact that the globally 
anisotropic Bianchi~V generalisations of the 
compact isotropic and homogeneous FLRW $k=-1$ models
prohibit the closure of their spatial sections,
thereby destroying their mixing capability.
This may be important for mixing scenarios
based on compact $k=-1$ models, if the anisotropies 
are global, since small anisotropies 
are bound to be present in the real universe.

What remains to be done is to find out the extent 
to which this type of fragility is also present
in anisotropic {\em and} inhomogeneous generalisations 
of the FLRW $k=-1$ models.

Finally, we should stress again that our results do not forbid 
the universe to be {\em locally} anisotropic-and-inhomogeneous
and have compact spatial section with the topology 
corresponding to a compact hyperbolic space.
They do, however, give an example of topological
fragility that might occur in other settings.
\vspace{3mm}

%%%%%%%%%%%%%%%%%%%%%%%%%%%%%%%%%%%%%%%%%%%%%%%%%%%%%%%%%%
\section*{Acknowledgements} 
%%%%%%%%%%%%%%%%%%%%%%%%%%%%%%%%%%%%%%%%%%%%%%%%%%%%%%%%%%

RT would like to thank CNPq for the financial support
which made his visit to CBPF possible and the colleagues 
in the Department of Relativity and Particles (DRP/CBPF)
for their warm hospitality. RT wishes to thank Henk van Elst
for helpful discussions.
He also benefited from SERC UK Grant No. H09454. 

MR and AT are also grateful to CNPq for
financial support.

%%%%%%%%%%%%%%%%%%%%%%%%%%%%%%%%%%%%%%%%%%%%%%%%%%%%%%%

%%%%%%%%%%%%%%%%%%%%%%%%%%%%%%%%%%%%%%%%%%%%%%%%%

\end{document}